\documentclass[default,iicol]{sn-jnl}

\usepackage{graphicx}%
\usepackage{multirow}%
\usepackage{amsmath,amssymb,amsfonts}%
\usepackage{amsthm}%
\usepackage{mathrsfs}%
\usepackage[title]{appendix}%
\usepackage{xcolor}%
\usepackage{textcomp}%
\usepackage{manyfoot}%
\usepackage{booktabs}%
\usepackage{algorithm}%
\usepackage{algorithmicx}%
\usepackage{algpseudocode}%
\usepackage{listings}%
\usepackage{mathtools} 

\usepackage{soul,xcolor}

\usepackage{hyperref} 
\hypersetup{
	colorlinks=true,
	linkcolor=blue,
	citecolor=blue,
	urlcolor=blue
}
\begin{document}

\setstcolor{red}

\title[Article Title]{Laplace's first law of errors applied to diffusive motion}
\author[1,2,*]{\fnm{Omer} \sur{Hamdi}}
\author[1,**]{\fnm{Stanislav} \sur{Burov} }
\author[1,2,***]{\fnm{Eli} \sur{Barkai}}
\affil[1]{Physics Department, \orgname{Bar-Ilan University}, \city{Ramat-Gan}, \postcode{52900}, \country{Israel}}
\affil[2]{\orgdiv{Institute of Nanotechnology and Advanced Materials}, \orgname{Bar-Ilan University}, \city{Ramat-Gan}, \postcode{52900}, \country{Israel}}

\vspace{-0.5in} 

\abstract{
In biological, glassy, and active systems, various tracers exhibit Laplace-like, i.e., exponential, spreading of the diffusing packet of particles. The limitations of the central limit theorem in fully capturing the behaviors of such diffusive processes, especially in the tails, have been studied using the continuous time random walk model.
For cases when the jump length distribution is super-exponential, e.g., a  Gaussian, we use large deviations theory and relate it to the appearance of exponential tails.
When the jump length distribution is sub-exponential the packet of spreading particles is described by the big jump principle.
We demonstrate the applicability of our approach for finite time, indicating that rare events and the asymptotics of the large deviations rate function can be sampled for large length scales within a reasonably short measurement time.

}

\maketitle

\section{Introduction}
\let\thefootnote\relax\footnotetext{Invited paper for the topical issue in the European Physical Journal B: New Trends in Statistical Physics of Complex Systems: Theoretical and Experimental Approaches.}
 Laplace's two laws of error are milestones in statistics.
The first was published in $1774$ \cite{laplace1774memoire} and states that the frequency of an error
could be expressed as an exponential of the magnitude of the error, in
 absolute value. The second law of errors, from $1778$ \cite{laplace1781memoire}, states that the  
frequency of the error is an exponential of a quadratic function
of the error.  In the context of diffusing particles in one dimension,
and for a packet starting at the origin, the probability density of the particles is $P(x,t)$.
The first law states $P(x,t) \propto \exp( - \mbox{const}\ |x|)$ and the 
second is the more familiar law, $P(x,t) \propto \exp( - \mbox{const}\ x^2)$.
In $1923$ Wilson \cite{wilson1923first} discussed some of the history of the problem. He noted that
the second law is typically called the normal or Gauss law; however,
despite Gauss's well-known precocity, he 
 probably did not make this discovery before he was two years old$^\dagger$ \footnote{$^\dagger$Gauss utilized the least squares method of error estimation for the discovery of the lost dwarf planet, Ceres. \cite{Gauss_discovery,609c561e-7c2f-3509-b332-91ef2a0a3670}.}. Indeed, it was Laplace who promoted the central
limit theorem, putting a firm mathematical basis for the second law. 
The normal distribution has since been used in all fields of science, while the first law, as far as we know,  was
not in the spotlight for an extended period. 

However, using single-molecule tracking data and computer-generated trajectories of a variety of tracers diffusing in disordered media, in more recent years, the first law experienced a revival, as it was used to fit a large body of 
experimental data \cite{pnas.2216497120,LAMPO2017532,C0SM00925C,PhysRevE.86.020901,Soares_e_Silva_2014,Weeks2000627,PhysRevLett.103.198103,STYLIANIDOU20142684,wang2009anomalous,wang2012brownian,rusciano2022fickian,PhysRevResearch.2.022020,guan2014even,PhysRevLett.126.158003,pastore2022model,waigh2023heterogeneous,chaudhuri2007universal,aaberg2021glass,miotto2021length}. 
These observations are related to the diffusing diffusivity models \cite{PhysRevLett.113.098302,yamamoto2021universal,waigh2023heterogeneous,e23020231} and a phenomena known as Brownian yet non-Gaussian diffusion \cite{wang2009anomalous,PhysRevX.7.021002,wang2012brownian,baldovin2019polymerization,Nampoothiri_2022,PhysRevE.104.L062501}, and Fickian yet non-Gaussian diffusion \cite{rusciano2022fickian,PhysRevResearch.2.022020,guan2014even,PhysRevLett.126.158003,pastore2022model}.
By now, the exponential decay of $P(x,t) \propto \exp(- \mbox{const}\ |x|) $ is well documented. In some cases, only the large $x$ limit is described by this law, while in others, the Laplace law, as a fitting procedure, holds for all $x$ \cite{wang2009anomalous}. A large body of phenomenological models was used to describe these behaviors, for example, by assuming that the diffusion constant $D(t)$ is a stochastic process.

Chaudhuri, Berthier, and Kob \cite{chaudhuri2007universal} analyzed four systems, focusing on particle displacements near glass and jamming transitions, highlighting behaviors like sticking (caging) and rapid jumps between basins.
This type of dynamics is described by the continuous time random walk (CTRW) \cite{montroll1965random,BARKAI200213,weiss1983random,weiss1994aspects,aghion2018asymptotic,kutner2017continuous,metzler2000random,Shafir_2022,barkai2000continuous,monthus1996models,masoliver2003continuous,klafter2011first,Vitali_2022,PhysRevE.109.014130}, which will also be analyzed in this paper. Using specific waiting time distributions, they showed how the basic CTRW model can be used to predict exponential tails, in accordance with Laplace's first law.
This theory was later advanced by Wang, Burov, and Barkai \cite{barkai2020packets,wang2020large}, showing that it holds for very general settings.
They have shown that, for any distribution of waiting times, which is analytical for short waiting times, and for any jump
length distribution, which decays faster than exponential, Laplace's first law
holds at the tails of $P(x,t)$. The two above-mentioned conditions are expected to hold in many systems. The analysis was based on large deviations \cite{TOUCHETTE20091,PhysRevLett.102.060601,PhysRevLett.113.078101,PhysRevLett.113.120601,PhysRevLett.121.090602,Derrida_2007,du2023dynamical,PhysRevE.103.042116} arguments, and a saddle point approximation \cite{lugannani1980saddle}, namely to put things in a historical context,
the technique used is an extension of the Laplace method for solving integrals \cite{laplace1774memoire}.

Schramma et al. \cite{pnas.2216497120} discovered that chloroplasts, which are components of plant cells, adapt to dim light by moving in a manner closely resembling the caged dynamics observed in supercooled liquids or colloidal suspensions near the glass transition. This movement demonstrates exponential tails, fitting both the CTRW and the diffusing diffusivity models.

The connection between Laplace diffusion and CTRW in polymer nanocomposites was further explored by Hu et al \cite{hu2023triggering}. They demonstrated the ability to manipulate the emergence of exponential tails in contrast to Gaussian diffusion by modifying the strength of the disorder.

In this manuscript, we first provide a more detailed analysis of CTRW dynamics using three tools. We will then conclude the paper with a broader perspective on the Laplace-like behavior, i.e., exponential decay of spreading particles.

We start with a study of the CTRW model for the case when the jump length distribution is either sub or super-exponential. In the former case large deviations theory holds, while for the latter the big jump principle is valid (see below). 
This transition is related to the fact that for sub-exponential jump length distribution, the cumulant generating function diverges, and thus the standard 
Laplace-Cramers-Daniles tool of saddle point approximation
of large deviations theory does not hold. 
We show how
the big-jump principle  
is related to an extension of the Laplace method of solving integrals. 
In Laplace's method, 
close to the saddle point,  an analytic i.e., quadratic function is used, and hence, eventually, a Gaussian
integral is computed, while the original integral is non-Gaussian. 
For the sub-exponential distribution of jump lengths, we find a similar
 extremum,
but with non-analytical features. Namely, the quadratic expansion close
to the extremum
is invalid.
Finally, we present the Edgeworth expansion to approximate the CTRW probability density \cite{rosenblatt1956central}.
This approximation provides corrections for the central limit theorem (CLT).
It deals with a long time limit, and not very large $x$, while the large deviations theory and the big-jump principle cover the behavior of the tails of the density.
 To study these effects, we constructed a numerical tool, to sample finite time and finite $x$ propagators $P(x,t)$ using CTRW.

\section{ Model}
In the CTRW model, waiting times between jump events are independent identically distributed (IID) random variables with a probability density function (PDF) $\psi(\tau)$. The process starts at time $t=0$, and then the particle waits at its initial position
$x=0$, for a random time $\tau$ drawn from $\psi(\tau)$. At time $t=\tau$, the random walker jumps to a new position, the duration of the jump being negligible.
The jump lengths are also IID random variables, with $f(\chi)$ as the PDF. 
The process is then renewed, namely, a second waiting time is drawn, followed by a spatial jump, etc. 
The goal is to find the PDF $P(x,t)$ of finding the particle at $x$ at time $t$.  
The CTRW process is called a semi-Markovian process
and the statistics for the number of jumps can be 
analyzed utilizing the renewal theory \cite{cox1962renewal,burov2020limit}.

We focus on exponentially distributed waiting times, $\psi(\tau)=
\exp( - \tau)$, with the mean time between jumps set to unity.
It follows that the number of jumps in the time interval $(0,t)$ is described
by Poisson statistics.
The jump length PDF $f\left( \chi \right)$ is:
\begin{equation}\label{eq01}
f\left( \chi \right) =  \widetilde{N} \exp\left( - \alpha^\beta |\chi|^\beta\right).
\end{equation}
The mean jump length is zero, and the symmetry of $f\left( \chi \right)$ sets the symmetry of $P(x,t)$, i.e.,  $P(x,t)=P(-x,t)$. The variance of the jump length is set to be $1$.
Here we have $\alpha = \sqrt{ \Gamma(3/\beta)/\Gamma(1/\beta)}$, and
the normalization $\widetilde{N} = \beta \sqrt{\Gamma(3/\beta)}/2 \Gamma^{3/2} (1/\beta)$, with $\Gamma(\dots)$ being the Gamma function. 
The exponent $\beta>0$ is an important parameter in this study.
For $\beta<1$ $(\beta>1)$ we have sub (or super) exponential decay
of the jump length distribution. The case $\beta=2$ corresponds to Gaussian 
statistics for the jump length PDF.

 The probability of jumping $n$ times during time $t$ is $t^n \exp(-t)/n!$ hence
\begin{equation}\label{eq:P(x,t)_summation}
P(x,t) = \sum_{n=0} ^\infty \frac{{\rm e}^{-t} t^n}{n!} \phi(x|n),
\end{equation}
$\phi(x|n)$ is the PDF of finding the particle at $x$, conditioned
on performing exactly $n$ jumps. 
In the previous equation, the summation is performed over all the possible number of jumps that can occur during the process.
Using the fact that jump lengths
are IID random variables, the characteristic function of $\phi(x|n)$
\begin{equation}
\int_{-\infty} ^\infty {\rm e}^{ i k x} \phi(x|n) {\rm d} x =
\langle {\rm e}^{ i k (\chi_1 + ... \chi_n) }\rangle= \widetilde{f}^n(k),
\label{eq:tilde{f}^n}
\end{equation}
is expressed in terms of the Fourier transform \cite{bochner1949fourier} of $f(\chi)$,
namely $ \widetilde{f}(k) = \int_{-\infty} ^\infty \exp( i k \chi) f(\chi) {\rm d} \chi$. 
The Fourier transform of $P(x,t)$, i.e., $\widetilde{P}(k,t)$, is obtained from
Eq.(\ref{eq:P(x,t)_summation}) and Eq.(\ref{eq:tilde{f}^n})
\begin{equation}\label{eq:P(k,t)}
\widetilde{P} (k, t)= \exp\left[ - t \left(1 - \widetilde{f}(k) \right) \right].
\end{equation}
The goal is to find the inverse Fourier transform of the expression above.
The CLT is valid in the limit $ t \to \infty$ and
$x \to \infty$ while the 
ratio 
$x/\sqrt{t}$ is finite. 
We utilize the expansion of $ \widetilde{f}(k)$ 
around $k\to 0$. Since the jump length distribution is an even function, $ \widetilde{f}(k) \sim 1 - \sigma^2 k^2 / 2$, and according to our notation $\sigma^2=1$. Therefore Eq.  (\ref{eq:P(k,t)}) yields
\begin{equation}\label{eq:P(x,t)_LDT_CLT_LIMIT}
P(x,t) \sim \frac{ \exp\left( - \frac{ x^2} {2 t}  \right )}{ \sqrt{ 2 \pi t} }.
\end{equation}
Namely, the central limit theorem holds.

\section{Large Deviations \texorpdfstring{$\beta\ > \ 1$}{beta > 1}}
Capturing the essence of the limiting scaling law in Eq. (\ref{eq:P(x,t)_LDT_CLT_LIMIT}) is but one aspect of the large deviations Theory. This theory also delves into analyzing a different yet significant limit.
The inverse Fourier transform of $\widetilde{P} (k, t)$ in Eq. (\ref{eq:P(k,t)}), while changing variable $i k =u$, reads
\begin{eqnarray}\label{eq:P(x,t)_InvFourier}
P(x,t) &=& \frac{1}{2 \pi i} \int_{- i \infty} ^{i \infty} \exp{\left[ - x K(u) \right]  }  {\rm d} u,
\end{eqnarray}
where
\begin{eqnarray}
K(u) &=& u + \frac{1}{q} [ 1 - \hat{f}(u)],
\end{eqnarray}
$q \equiv x/t$, and $\hat{f}(u) = \int_{-\infty} ^\infty \exp( u \chi) f(\chi) {\rm d} \chi$ is the moment generating function. 
Note that $\hat{f}(u)$ diverges for $\beta<1$, leading to the divergence of the integral in Eq. (\ref{eq:P(x,t)_InvFourier}). Therefore, in this section, we focus on super-exponential PDFs, i.e., $\beta>1$. 

Large deviations theory is based on the saddle point approximation and considers the case of large $x$ limit of $P(x,t)$ in Eq. (\ref{eq:P(x,t)_InvFourier}).
Strictly speaking, here the scaling is $x\to \infty$ and $t\to \infty$ while the ratio  $q=x/t$ is kept finite. Nevertheless, we show that the approximate solution for $P(x,t)$ works reasonably well for large $x$ while $t$ is kept finite.

The problem of finding $P(x,t)$ is solved by using the following steps: First, find
the $u$ that satisfies $K'(u)=0$ and term this $u$ as $u_0$.
Then, expand $K(u)$ in Eq. (\ref{eq:P(x,t)_InvFourier}), in the vicinity of $u_0$,
up to a quadratic term. The obtained Gaussian integral yields
\begin{equation}\label{eq:P(x,t)_LDT_K(u_0)}
P(x,t) \sim \frac{\exp\left[ - x K(u_0)\right]}{ \sqrt{ 2 \pi |x K''(u_0)|} }.
\end{equation}
The approximation for large x can be obtained through numerical methods.
Namely, we numerically find $u$ that satisfy $K'(u)=0$, and insert it into Eq. (\ref{eq:P(x,t)_LDT_K(u_0)}). 
We call this solution method the numerical large deviations method (Numerical-LDT).
Below, we discuss how to obtain the exact form of $P(x,t)$ and compare it to the solution obtained via Numerical-LDT.

Using $K'(u)=0$, and the definition of $\hat{f}(u)$, the equation for $u_0$ reads
\begin{equation}
1 - \frac{1}{q} \int_{-\infty} ^\infty \chi f( \chi) \exp( u_0 \chi) {\rm d} \chi =0.
\label{eq:exact_u_0_forumula}
\end{equation}
We treat this equation in two limits, small and large $q = x/t$.
For small $q$ the value of integral in Eq. (\ref{eq:exact_u_0_forumula}), must be small.
Since the integral in Eq.~\eqref{eq:exact_u_0_forumula} is a growing function of $u$, the limit of small $q$ corresponds to small values of $u$. Expanding $\exp( u \chi)$ and using
the fact that the variance of jump lengths equals $1$ we find
$u_0\sim q \ll 1$, and $K(u_0)\sim u_0 + \frac{1}{q} [1 - \hat{f}(u_0)]
\sim q/ 2$. It then follows that for $q\ll 1$, $P(x,t)\propto \exp[ - x^2/(2 t)]$.
This is the standard prediction of the CLT, as delineated in Eq. (\ref{eq:P(x,t)_LDT_CLT_LIMIT}),
hence the more interesting case is $q\gg 1$. 

We use Eq. (\ref{eq:exact_u_0_forumula}) in the large $q$ limit, considering large $u$, to find $u_0$. For that aim, we need to
find $\hat{f}(u)$ when $u$ is large.
With a change of variable $\chi= u^{\frac{1}{\beta-1}} \xi$, we obtain
\begin{equation}
\hat{f}(u) = \widetilde{N} u^{ \frac{1}{\beta -1}}
\int_{-\infty} ^\infty \exp\left[ u^{ \frac{\beta}{ \beta -1}} \left( \xi - \alpha^\beta |\xi|^\beta\right) \right] {\rm d} \xi.
\end{equation}
Since $u$ is large this integral is solved using Laplace's method
\begin{equation}
\hat{f}(u) \sim \widetilde{N} C_2 u^{ \frac{1 }{ 2} \frac{ 2 - \beta }{ \beta -1} }
\exp\left( C_1 u^{ \frac{\beta }{\beta-1}} \right),
\label{eq:hat{f}(u)}
\end{equation}
where $C_1= (\beta-1)(\beta \alpha)^{\frac{\beta}{1-\beta}}$, and $C_2 = \sqrt{2\pi/(\beta-1)(\beta\alpha^{\beta})^{\frac{1}{\beta-1}}}$. This expression is exact for the
Gaussian case $\beta=2$. Eq. (\ref{eq:exact_u_0_forumula}) for $u_0$ is written as $0= 1 - \frac{1}{q}  \hat{f'}(u)$, and using Eq. (\ref{eq:hat{f}(u)}) we find that $u_0$ satisfies
\begin{eqnarray}\label{eq:u_0 approx}
0 = 1-\frac{1}{q} \frac{\beta \widetilde{N} C_1 C_2}{\beta-1}u_0^{\frac{4-\beta}{2\beta-2}} \exp \left(C_1 u_0^{\frac{\beta}{\beta-1}}\right).
\end{eqnarray}
Note that in the case of $\beta = 4$, this approximation breaks down. For that reason, in this work, we focused on the range $1<\beta<4$ and the transition to $\beta<1$, where the big jump principle holds.
To solve Eq. (\ref{eq:u_0 approx}) we utilize the Lambert function $W(x)$ \cite{corless1996lambert} which satisfies $W(x) \exp\left[W(x)\right]=x$. The branch of the Lambert function that is relevant to our study is the well-documented principal branch $W_0 (x)$. Solving Eq.
(\ref{eq:u_0 approx}) we obtain
\begin{equation}
u_0 \sim C_3 W_0 \left[  C_4 q ^{ \frac{ 2 \beta  }{ 4-\beta}}
\right]^\frac{\beta-1 }{ \beta} .
\end{equation}  
For large values of the argument, the Lambert function is expressed in terms of logarithmic functions, i.e., $W_0 (z) \underset{z\to\infty}{\sim} \ln(z) - \ln(\ln(z))$,
 hence $u_0$ is growing with $x$, however very slowly. This holds true only
when $\beta<4$, while the opposite case demands further study. 
The constants are:
\begin{eqnarray}
     C_3= \beta \alpha \left( \frac{2\beta(\beta - 1)}{4 - \beta} \right)^{\frac{1 - \beta}{\beta}}\quad, \quad \nonumber\\ 
     C_4=\frac{2(\beta(\beta - 1))^{\frac{4}{4 - \beta}}}{4 - \beta} \left( \frac{\alpha^2}{\sqrt{2\pi} \widetilde{N}(\beta)} \right)^{\frac{2\beta}{4 - \beta}} .
\end{eqnarray}
Using $\exp[ a W(x)] = (x)^a / [ W(x)]^a$, we obtain $K(u_0)=u_0 + \frac{1}{q} [1 -\hat{f}(u_0)]$. Therefore, according to Eq. (\ref{eq:P(x,t)_LDT_K(u_0)}), we have that

%
\begin{equation}\label{eq:P(x,t)_ARF}
P(x,t) \sim \frac{\exp \left\{-t\left(\frac{|x|}{t}Z\left(\frac{|x|}{t} \right)+1\right) \right\}}{\sqrt{2\pi K^{''}(u_0)}},
\end{equation}
where 
\begin{equation}
    Z(q)=\frac{c_3 W_0\left[c_4 q^{\frac{2\beta}{4-\beta}}\right]-\alpha \left( \frac{2 \beta (\beta - 1)}{4 - \beta} \right)^{\frac{1}{\beta}}}{W_0\left[c_4 q^{\frac{2\beta}{4-\beta}}\right]^{\frac{1}{\beta}}}
\label{eq:z(q)}.
\end{equation}
Using $W_0(x)\sim \ln(x)$ we find
\begin{equation}
P(x,t) \sim \frac{\exp\left\{ - t\left[\kappa \frac{|x|}{t} \log\left(\frac{|x|}{t}\right)^{1-1/\beta} +1\right]\right\} }{\sqrt{2\pi K^{''}(u_0)}},
\label{eq:P(x,t)_lambert_as_log}
\end{equation}
where $\kappa = \beta \alpha/(\beta-1)^{1-1/\beta}$.
Thus, for large $x$, the packet is decaying according to Laplace's first law, with logarithmic corrections which are typically hard to detect
in experiments.
The result in Eq. (\ref{eq:P(x,t)_lambert_as_log}) was found in a more general setting in \cite{barkai2020packets}.
We also note a mistake in Eq. (9) in \cite{barkai2020packets} the later equation corresponds to Eq. (\ref{eq:z(q)}). Finally, we do not advocate the practical use of Eq. (\ref{eq:P(x,t)_lambert_as_log}) beyond an asymptotic result. The Lambert functions in Eq. 
(\ref{eq:z(q)}) are needed unless, $x/t$ is astronomically large.
To see this consider the asymptotic expansion $W_0(x)\sim \ln(x) - \ln(\ln(x))$
the second term is one percent of the first when $x \sim 10^{280}$. In other words, the leading term, in applications not sampling extremely rare events is insufficient.
We now show how the theory is applicable for finite times, using numerically exact $P(x,t)$.
\begin{figure}[ht]
\centering
\includegraphics[width=0.45\textwidth]{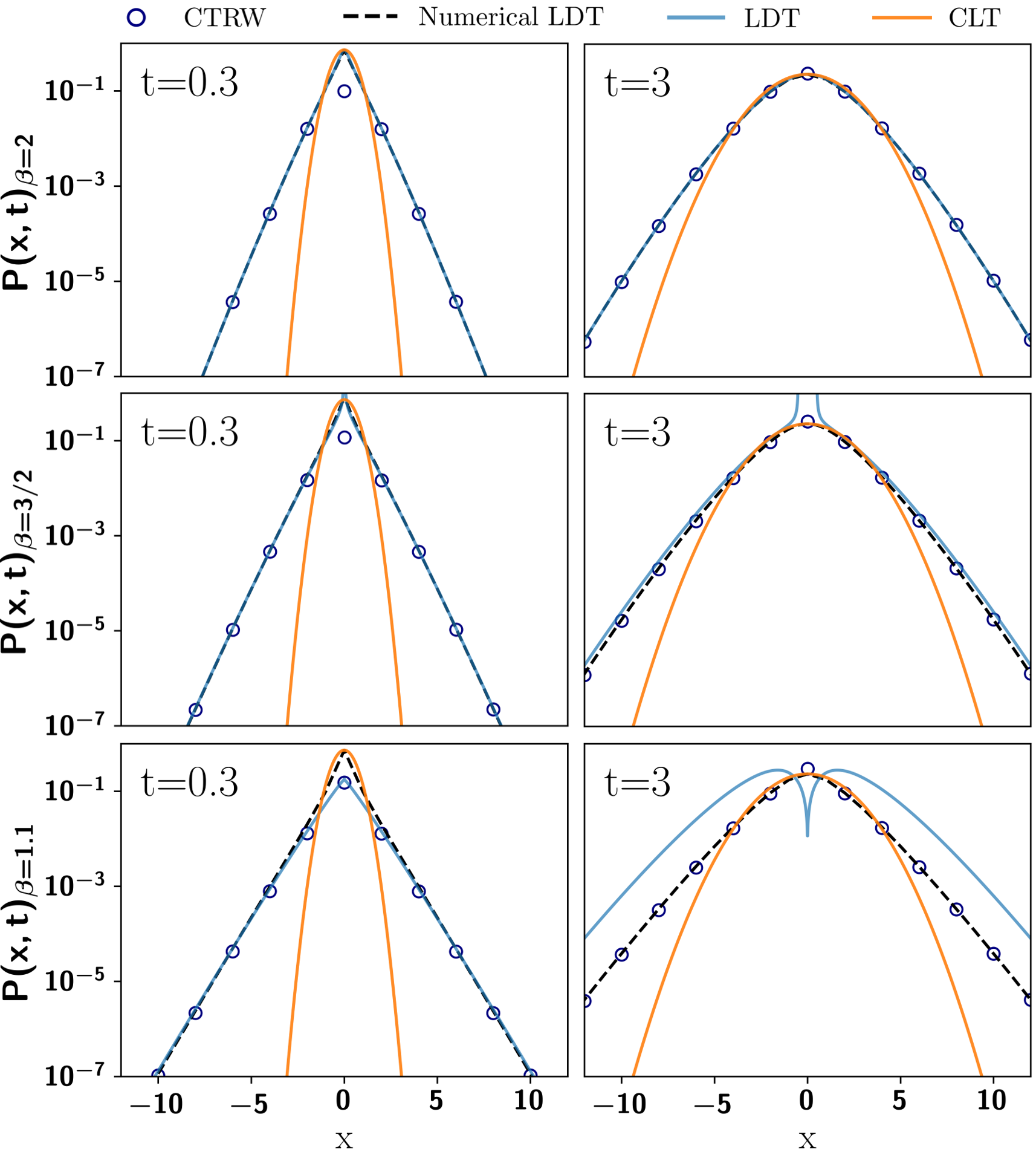}
\caption{Comparative analysis of the CTRW exact solution (circles), our approximation derived from Eq. (\ref{eq:P(x,t)_ARF}) (blue line), numerical LDT derived using Eq. (\ref{eq:P(x,t)_LDT_K(u_0)}) (dashed black), and CLT predictions (orange line), log-plotted for two time scales (0.3,3), and various $\beta$ values.
Notably, the nearly exponential tails are a dominant feature in both time scales, with an observable convergence to the CLT regime over time.
The numerical LDT shows remarkable fitting in the tails and also exhibits a strong performance in the central part as time progresses.
The approximation we propose in Eq.(\ref{eq:P(x,t)_ARF}) demonstrates a close match with the numerical data, however, deviations become more pronounced for large time values as $\beta \rightarrow 1$. This trend, indicative of slower convergence at smaller $\beta$ values, is further analyzed and discussed in the following section.}\label{Fig:P(x,t)}
\end{figure}

\subsection{CTRW Propagator}
Sampling the ``rare events" of the CTRW, particularly the tails of the PDF, poses significant challenges in trajectory simulations. To address this, we utilize the convolution theorem of the Fourier transform to compute $\phi(x|n)$ in Eq (\ref{eq:P(x,t)_summation}). More specifically, in Fourier space, the equation $\hat{\phi}(k|n) = \widetilde{f}^n(k)$ holds true. This leads to the recursive relation: $\phi(x|n) = f \ast f \dots \ast f = \phi(x|n-1)\ast f$, where $*$ is the convolution operator. For the case of $N=2$, this translates to $\phi(x|2)=f \ast f=\int_{-\infty}^{\infty} f(\chi')f(x-\chi')d\chi'$. This equation for $\phi(x|2)$ represents the integration over the probability of making a first step to a certain point $\chi'$ and then a subsequent step to $x$.

In Fig. \ref{Fig:P(x,t)}, we conduct a comparison between this numerically exact solution (CTRW), our derived approximation in Eq. (\ref{eq:P(x,t)_ARF}), and the numerical LDT calculated using Eq. (\ref{eq:P(x,t)_LDT_K(u_0)}), juxtaposed with the predictions of the CLT. For short time scales, our derived approximation and the numerical LDT demonstrate excellent agreement in the tails, where the CLT fails. As the time scale increases, the numerical LDT consistently aligns for all values of $x$ and $\beta$, whereas our derived approximation, though precise for $\beta=2$ (Gaussian), and closely fitting for $\beta = 3/2$, converges more gradually when $\beta=1.1$, nearing the critical transition for $\beta=1$ that was studied in \cite{singh2023universal}. For this case, when $\beta = 1$, exponential-like tails are still exhibited but are not described by Eq. (\ref{eq:P(x,t)_lambert_as_log}).

Upon further examination of Fig. \ref{Fig:P(x,t)}, it becomes apparent that for $t\gg 1$, the application of the CLT or the Edgeworth expansion is also viable, a topic we will discuss shortly.
\begin{figure}[ht]
\centering
\includegraphics[width=0.45\textwidth]{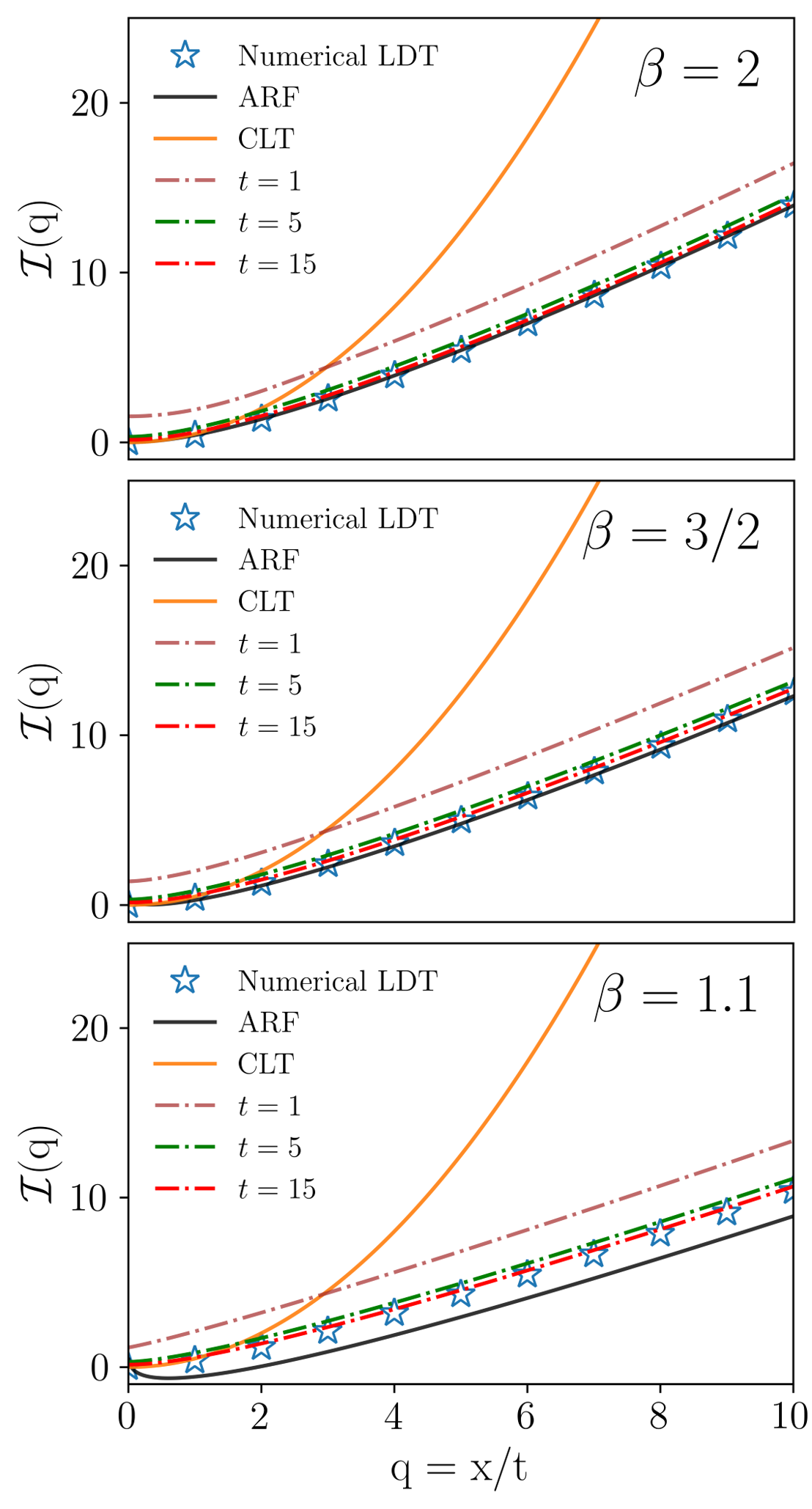}
\caption{Convergence of the CTRW results (dashed-dotted) for increasing time scales, $t=1, 5, 15$, towards the numerical LDT, as derived in Eq. (\ref{eq:P(x,t)_LDT_K(u_0)}) (blue stars). The theory demonstrates exceptional accuracy, with the CTRW aligning with our numerical LDT across all values of $\beta$ for long times. Additionally, the linear-like nature of the tails is observed, aligning with the expected Laplace's exponential tails. We also plot the asymptotical rate function (ARF) from Eq. (\ref{eq:LDT_rate_function_asymptotics}) (shown as a black line). Notably, for $\beta = 2$, the ARF aligns perfectly with the predictions, and for $\beta = 1.5$, we achieve a close fit. However, as we approach the regime of the big jump principle, for $\beta = 1.1$, convergence is slow, as we later describe.}\label{Fig:RateFunction}
\end{figure}

\subsection{Rate Function}
The rate function is an important concept in the large deviations literature as it holds the main characteristics of the propagator, describing both the typical and rare events. It is obtained in the limit of $t\rightarrow \infty$, but $x/t$ finite:
\begin{eqnarray}
    \mathcal{I}(x/t) = \lim_{t \to \infty} -\frac{1}{t}\ln{\left[P(x,t)\right]}.
\end{eqnarray}
In our model, we can extract the rate function from Eq. (\ref{eq:P(x,t)_LDT_K(u_0)}). Taking into consideration the fact that $u_0$ is a function of $q=x/t$, see Eq. (\ref{eq:exact_u_0_forumula}), i.e. $q$ is the scaling parameter of the rate function:
\begin{eqnarray}\label{eq:LDT_rate_function_asymptotics}
    \mathcal{I}(q) = q K(u_0) \approx
      \begin{dcases*} 
    q^2/2 & , $q\ll 1$ \\
    |q|Z(|q|)+1& , $q\gg 1$ .
  \end{dcases*}
\end{eqnarray}
The rate function in the small $q$ limit exhibits the quadratic CLT behavior, whereas the large $q$ behavior is what we call the asymptotic rate function (ARF). In Fig. \ref{Fig:RateFunction}, we demonstrate how our model effectively captures both the central part, corresponding to the CLT, and the Laplace(exponential)-like tails, corresponding to the large $q$ limit in Eq. (\ref{eq:LDT_rate_function_asymptotics}), as $Z(q)$ is a slowly varying function. It becomes evident that, as time progresses, the numerical solution (CTRW) converges to our numerical LDT rate function. Moreover, although the ARF theory is valid, as expected, we observe an issue of slow convergence for values of $\beta$ near 1.

\begin{figure}[ht]
\centering
\includegraphics[width=0.45\textwidth]{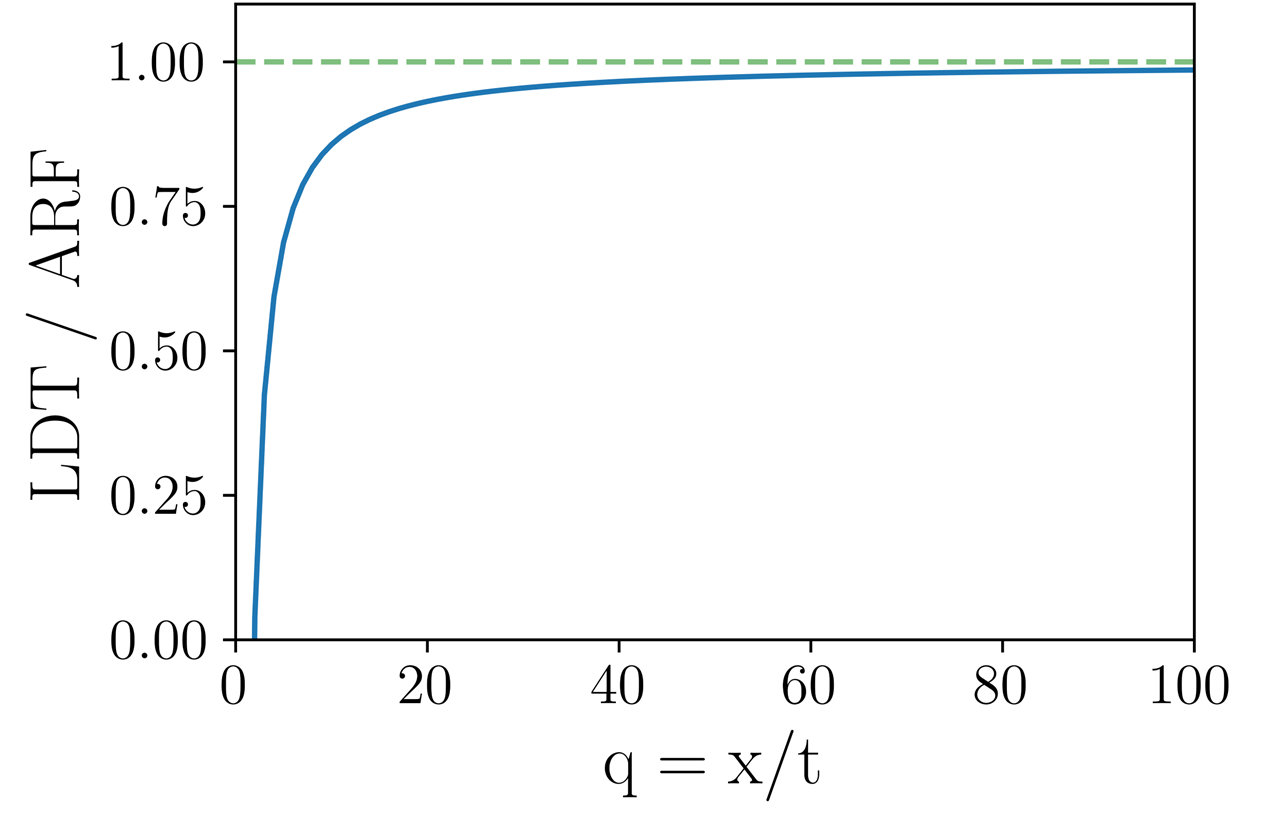}
\caption{Slow convergence of the asymptotical rate function (ARF) as formulated in Eq. (\ref{eq:LDT_rate_function_asymptotics}), to the numerical LDT obtained using Eq. (\ref{eq:P(x,t)_LDT_K(u_0)}), for the case $\beta = 1.1$. As we approach $\beta =1$ from above, the convergence of our approximation (ARF) becomes increasingly gradual. This slowing of convergence is attributed to the proximity to the transition point. At this critical juncture, the moment-generating function becomes non-existent, and the system transitions into the big jump principle (BJP) regime.}\label{Fig:SlowConv}
\end{figure}
\subsection{Slow Convergence of the ARF}
The issue of slow convergence of the ARF, as observed in Fig. \ref{Fig:RateFunction}, becomes increasingly pronounced as we mentioned before. More specifically, we focus on the scenario where our approximation, formulated in Eq. (\ref{eq:LDT_rate_function_asymptotics}), under the umbrella of LDT $(\beta>1)$, is nearing the critical transition point that occurs for $\beta = 1$. For this value of $\beta$, the moment-generating function ceases to exist and our approximation becomes invalid.

\begin{figure}[ht]
\centering
\includegraphics[width=0.45\textwidth]{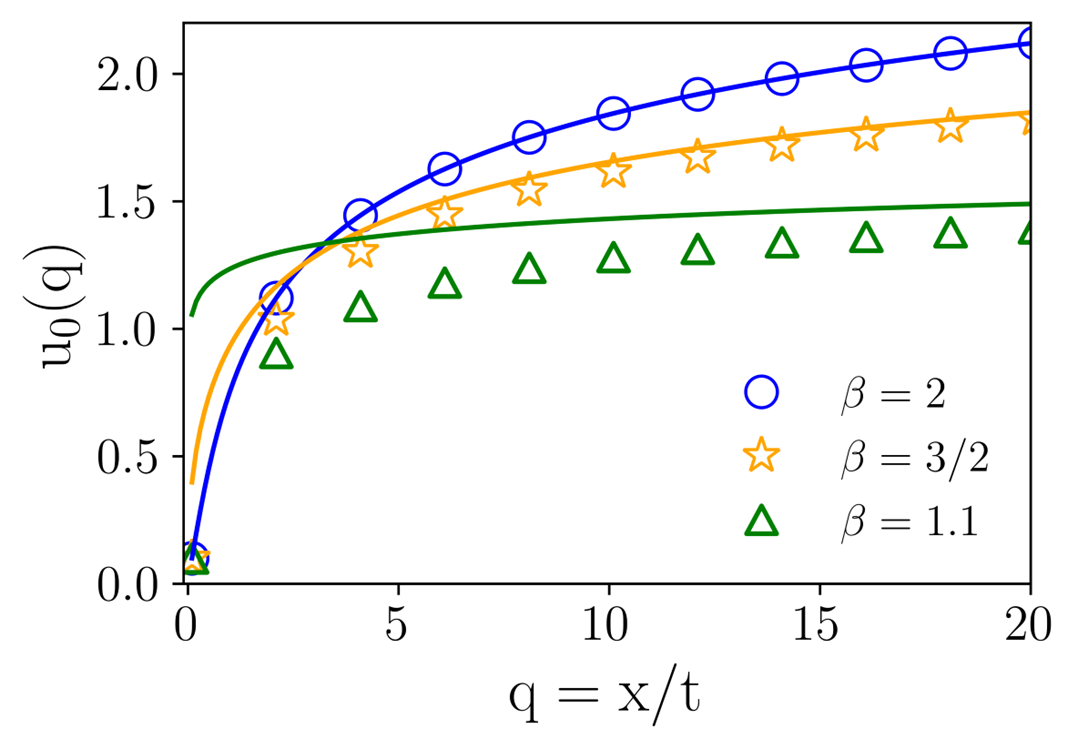}
\caption{Approximation of $u_0$ (colored lines), and numerically computed $u_0$ (corresponding colored symbols). While for $\beta = 2$ our approximation is exact, as $\beta \rightarrow 1$ the approximation works well only when $x\gg t$.}\label{Fig:u_0_SlowConvergence}
\end{figure}

In Fig. \ref{Fig:SlowConv}, we highlight how, for $\beta = 1.1$, the ARF converges slowly to the numerical LDT, as $q=x/t\rightarrow \infty$. This gradual convergence is a direct consequence of our approximation for $u_0$, and the numerical solution for $u_0$ eliminates this problem. In Fig. \ref{Fig:u_0_SlowConvergence} we demonstrate how our approximation of $u_0$ worsens as $\beta$ approaches 1, leading to the slow convergence of the rate function and the propagator in such cases.

\section{Edgeworth Expansion}
In the context of the summation of n IID random variables, the Edgeworth expansion \cite{edgeworth1905law,68a853b4-f1dd-3632-a4a2-c66e4433cd98,kendall1948advanced} provides corrections to the CLT.
As a reminder, we denote the PDF of finding the particle at $x$, conditioned
on performing $n$ jumps, as $\phi(x|n)$. The leading order of the Edgeworth expansion when $n$ is large, is:
\begin{equation}\label{eq:EdgeWorth_n}
\phi(x|n) \sim
\frac{ \exp( - x^2 / 2 n) }{ \sqrt{2 \pi n} }\left[ 1 + \frac{ \kappa_4 \mathrm{He}_4(x/\sqrt{n}) }{ 4!\ n} + \cdots  \right]
\end{equation}
where $\kappa_4$ is the fourth cumulant of $f(\chi)$, namely
$\kappa_4= m_4 - 3 \sigma^2$, and $m_4$ is the forth moment of $f(\chi)$. In our example
 $\sigma=1$ while $m_4=\Gamma(1/\beta) \Gamma(5/\beta) / \Gamma(3/\beta)^2$,  $\mathrm{He}_4(\xi) = \xi^4 - 6 \xi^2 + 3$.
In Eq. (\ref{eq:EdgeWorth_n}) the correction term is proportional to the Hermite polynomial. 

It is tempting to use the Edgeworth expansion in the summation presented in Eq. 
(\ref{eq:P(x,t)_summation}). However, if $\kappa_4=0$, namely when $f(\chi)$ is Gaussian corresponding to
$\beta=2$, the correction term vanishes. We introduce a modified
Edgeworth expansion, where $t$ is the large parameter.
We use the small $k$ expansion $ \widetilde{f} (k) = 1 - k^2 /2 + m_4 k^4 /4! + \cdots$ in Eq. (\ref{eq:P(k,t)})
\begin{equation}
\widetilde{P}(k,t) = \exp\left[ -\frac{ t k^2 }{ 2} + \frac{t m_4 k^4 }{ 4! } + \cdots\right].
\end{equation}
Further expanding
\begin{equation}
\widetilde{P} (k,t) = \exp\left( - \frac{t k^2 }{ 2} \right)\left[
1 + k^4 \frac{m_4 t }{ 4!} + \cdots \right].
\end{equation}
Performing the inverse Fourier transform we obtain
\begin{equation}\label{eq:P(x,t)_edgeworth}
P(x,t) =  \frac{\exp\left( - x^2/2 t \right)}{\sqrt{2 \pi t}} \left[ 1 + \frac{  m_4 }{ 4!\ t} \mathrm{He}_4 \left(\frac{x}{ \sqrt{t}}\right)+ \cdots \right], 
\end{equation}
Eq. (\ref{eq:P(x,t)_edgeworth}) has the same structure as the original expansion in Eq. (\ref{eq:EdgeWorth_n}), where we replaced large $n$ with large $t$. The difference is that now the correction is proportional to the fourth moment of the distribution $f(\chi)$ and not to its cumulant. The former is always non-negative while the latter can be negative zero or positive. Therefore, we see that the fluctuations in the number of jumps
have a significant effect.  

\begin{figure}[ht]
\centering
\includegraphics[width=0.45\textwidth]{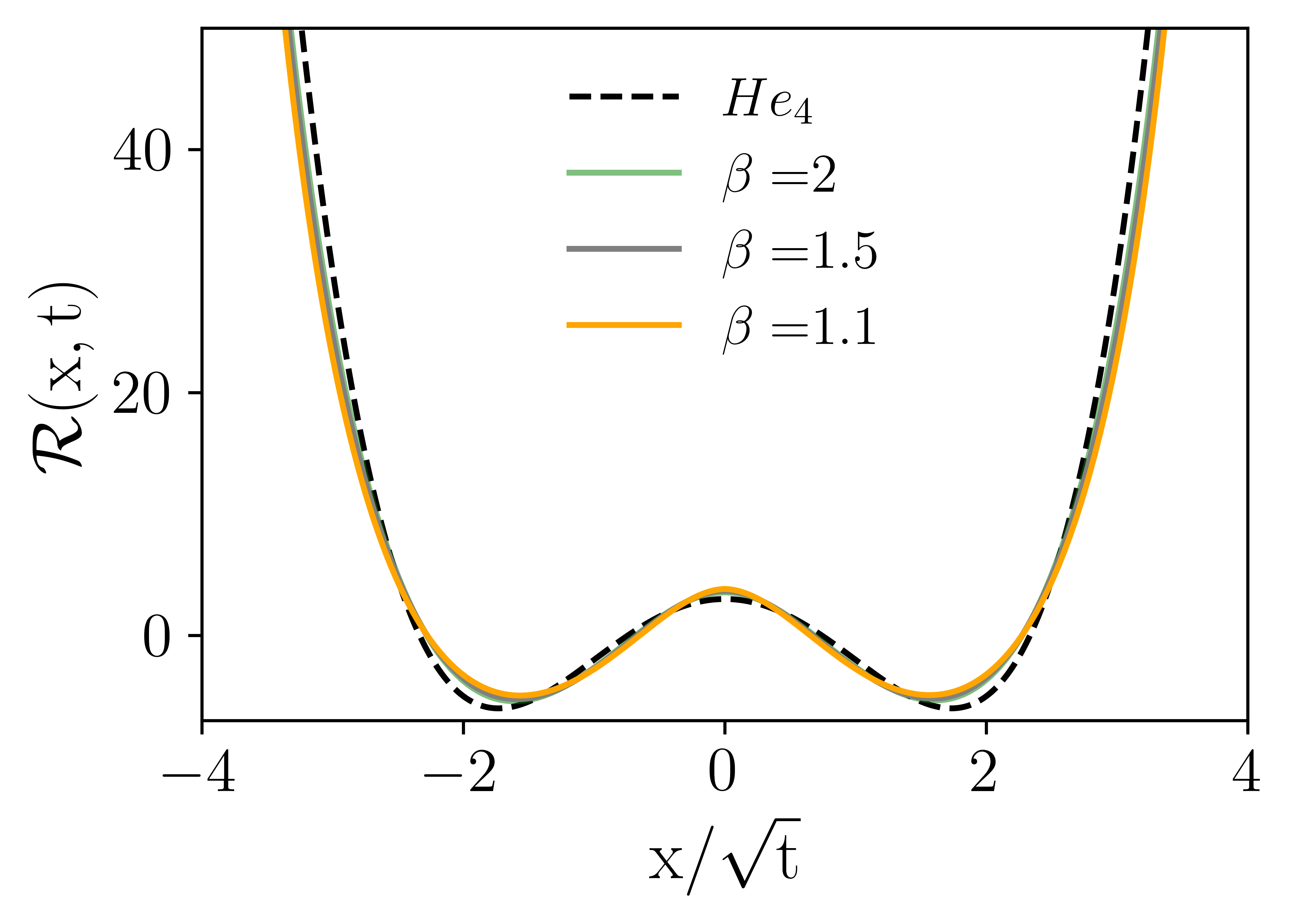}
\caption{We showcase the leading order of the Edgeworth expansion by plotting, for $t=10$ the rearranged result from the numerical CTRW as outlined in Eq.(\ref{eq:rearranged_hermite}) (colored lines). This is compared against the fourth Hermite polynomial $\mathrm{He}_4(x/\sqrt{t}) = (x/\sqrt{t})^4 - 6 (x/\sqrt{t})^2 + 3$. The excellent agreement is observed regardless of the value of $\beta$, thereby confirmming the validity of this expansion for large time scales.}\label{Fig:4Th_Hermite_polynomial}
\end{figure}

The contribution of the Hermite polynomials within our model can be demonstrated through a straightforward rearrangement of $P(x,t)$ from Eq. (\ref{eq:P(x,t)_edgeworth}), denoted as $\mathcal{R}(x,t)$. By defining the central limit theorem component as $\mathrm{CLT}=\exp\left( - x^2/2 t \right)/\sqrt{2 \pi t}$, we obtain the following form:
\begin{eqnarray}\label{eq:rearranged_hermite}
    \mathcal{R}(x,t)=\frac{4! t}{m_4}\frac{P(x,t) -CLT}{CLT},
\end{eqnarray}
where, at sufficiently long times, this expression approaches the fourth Hermite polynomial. This is shown in Fig. \ref{Fig:4Th_Hermite_polynomial} for different values of $\beta$, where $\mathcal{R}(x,t)$ is plotted versus the scaled variable $x/\sqrt{t}$.

The Edgeworth expansion uses the diffusive scaling $x/\sqrt{t}$ while the large deviations theory relies on the scaled variable $x/t$. Both contribute to the leading order of the central limit theorem when $x/t\ll 1$.
Note that when using the Edgeworth approach in the limit $x/\sqrt{t} \gg 1$, we have from the Hermite function a correction term to the Gaussian $P(x,t)$ that can be written as 
$t m_4 (x^4 /t^4)$. This results in the same scaling $q=x/t$ as found in the rate
function $P(x,t) \propto \exp [ - t {\cal{I}} (x/t)]$. 
Thus, as expected, the 
 Edgeworth expansion and large deviations theory are not conflicting.
They consider different limits of the problem. 
The large deviations theory is valid for $\beta>1$ as it demands
finite cumulants.
The convergence properties of the Edgeworth expansion and their $\beta$ dependence are left for future work.
\section{Big Jump Principle}
\subsection{Sum of IID Random Variables}
The big jump principle (BJP) \cite{vezzani2019single, vezzani2020rare, chistyakov1964theorem, H_ll_2021, wang2019transport, burioni2020rare, vezzani2023fast} describes how, in a heavy-tailed process with a sub-exponential distribution of jump lengths, a single ``big" jump among a series of $n$ IID random variables can dominate the asymptotic characteristics of the process. This principle connects the sum $x=\sum_{i=1}^n \chi_i$ to the largest displacement $\chi_{\text{max}} = \max {\chi_1,\dots,\chi_n}$ \cite{chistyakov1964theorem}. Specifically, this principle provides the asymptotic equality:
\begin{eqnarray}\label{eq:BJP_P}
    \int_{\chi_{\text{large}}}^{\infty}\phi(x|n)dx &=& \text{Prob}(x>\chi_{\text{large}}) = \nonumber \\
    && \text{Prob}(\chi_{\text{max}}>\chi_{\text{large}}),
\end{eqnarray}
in the limit of $\chi_{\text{large}} \rightarrow \infty$. To further elucidate, we calculate:
\begin{eqnarray}
    \text{Prob}&&\left(\chi_{\text{max}}>\chi_{\text{large}}\right) = 1-\text{Prob}\left(\chi_i<\chi_{\text{large}}\right)^n \nonumber\\
    &&\quad = 1-\left[1-\int_{\chi_{\text{large}}}^{\infty}f(\chi)d\chi\right]^n.
\end{eqnarray}
Taking the derivative with respect to $\chi_{\text{large}}$, we obtain the PDF of $x$. Then, by approximating $\int_{\chi_{\text{large}}}^{\infty}f(\chi)d\chi$ as approximately zero, which is a leading-order approximation, we get the asymptotic result \cite{foss2011introduction}:
\begin{eqnarray}\label{eq:BJP_leading_term}
    \phi(x|n)_{\beta<1} \approx n f(x).
\end{eqnarray}
It is important to note that, in contrast with the CLT, which is applicable primarily when $n$ is significantly large, Eq. (\ref{eq:BJP_leading_term}) is valid for all values of $n$, including instances when $n=2$.
We will discuss the BJP for CTRW later, but for now, our focus is on $\phi(x|n)$, specifically in the case where the number of jumps, $n$, equals 2.
\begin{figure}[ht]
\centering
\includegraphics[width=0.45\textwidth]{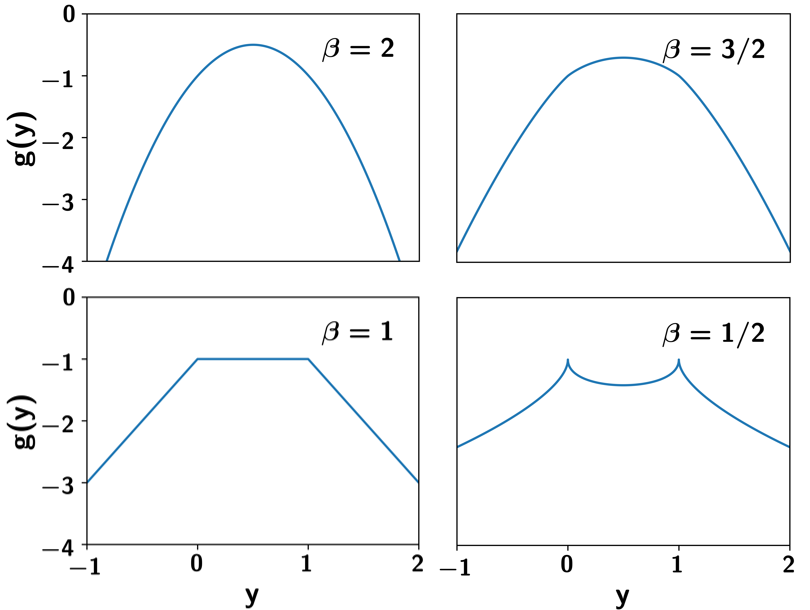}
\caption{The graph of $g(y)=-|1-y|^\beta -|y|^\beta$, for different values of $\beta$. It is evident that for $\beta >1$,  the function is smooth with a single global maximum at $y_0=1/2$. For $\beta =1$, the function is piecewise linear, and for $\beta <1$, it exhibits two non-analytical cusps at $y=0$, and $y=1$. This example serves as a foundational demonstration of the transition from BJP $(\beta <1)$ to LDT $(\beta > 1)$, while $\beta=1$ is a special transition point \cite{singh2023universal}.}\label{Fig:Transition_Two_Peaks}
\end{figure}
\subsection{Big Jump Principle for $n = 2$}
An understanding of the big jump principle is best demonstrated through the simple yet comprehensive scenario of two jumps. Employing the convolution theorem as seen earlier; $\phi(x|2)=f \ast f = \widetilde{N}^2\int_{-\infty}^{\infty} \exp\left(-\alpha^\beta |\chi'|^\beta-\alpha^\beta |x-\chi'|^\beta\right) d\chi'$, we introduce the scaling variable $y=\frac{\chi'}{x}$, and obtain the following equation:
\begin{eqnarray}\label{eq:phi(x|2)_exact}
    \phi(x|2) = \widetilde{N}^2 |x| \int_{-\infty}^{\infty} {\rm e}^{\alpha^\beta|x|^\beta g(y)}dy,
\end{eqnarray}
where we define $g(y) \equiv -|1-y|^\beta -|y|^\beta$.
In Fig. \ref{Fig:Transition_Two_Peaks} we demonstrate the transition in the analytical properties of $g(y)$. For $\beta>1$ where $f(\chi)$ is super-exponential, $g(y)$ is a smooth function with a single global maximum at $y_0=1/2$, suitable for the Laplace method of solving integrals.
Conversely, as shown in Fig. \ref{Fig:Transition_Two_Peaks}, for $\beta<1$, $g(y)$ displays a pair of cusps, and two distinct maxima are observed. We proceed to examine these cases. Beginning with the case of $\beta>1$, one can approximate $g(y) \approx g(y_0) + \frac{1}{2}g^{''}(y_0)(y-y_0)^2$, facilitating a direct Gaussian integration:
\begin{eqnarray}
    \phi(x|2)_{\beta>1} \approx  \widetilde{N}^2 |x| \sqrt{\frac{\pi 2^{\beta-2}}{\beta (\beta -1)\alpha^\beta |x|^\beta}}{\rm e}^{-\frac{1}{2}^{\beta-1}\alpha^\beta |x|^\beta}. \nonumber\\
\end{eqnarray}
This result diverges as $\beta$ approaches 1. The critical case of $\beta =1$ marks a transition, where the cumulant generating function of $\phi(x|n)$, exists only for $\beta>1$, and a piecewise linear function is displayed. At $\beta=1$, we can exactly solve Eq. (\ref{eq:phi(x|2)_exact}), yielding: $\phi(x|2)= \frac{\sqrt{2} + 2 | x | }{4} {\rm e}^{-\sqrt{2} | x |} = \left(\frac{1+\sqrt{2}|x|}{2}\right)f(x)$, which is not described by either the BJP or the LDT.

However, for $\beta<1$, where $f(\chi)$ is sub-exponential, $g(y)$ exhibits two non-analytical cusps, and a slightly modified version of the Laplace method of solving integrals is needed.
Due to the symmetry of $g(y)$, it is easy to show that the main contributions from the two maxima to the integration are identical. Opting for the left cusp, around $y=0$, we approximate $g(y)  \approx -1-|y|^\beta$, and insert it into Eq. (\ref{eq:phi(x|2)_exact}), to derive: $\phi(x|2) \approx  \widetilde{N}^2 |x| \int_{-\infty}^{\infty} {\rm e}^{-\alpha^\beta|x|^\beta \left(1+|y|^\beta\right)}dy= 2 f(x)$, aligning with the BJP.

\begin{figure}[ht]
\centering
\includegraphics[width=0.45\textwidth]{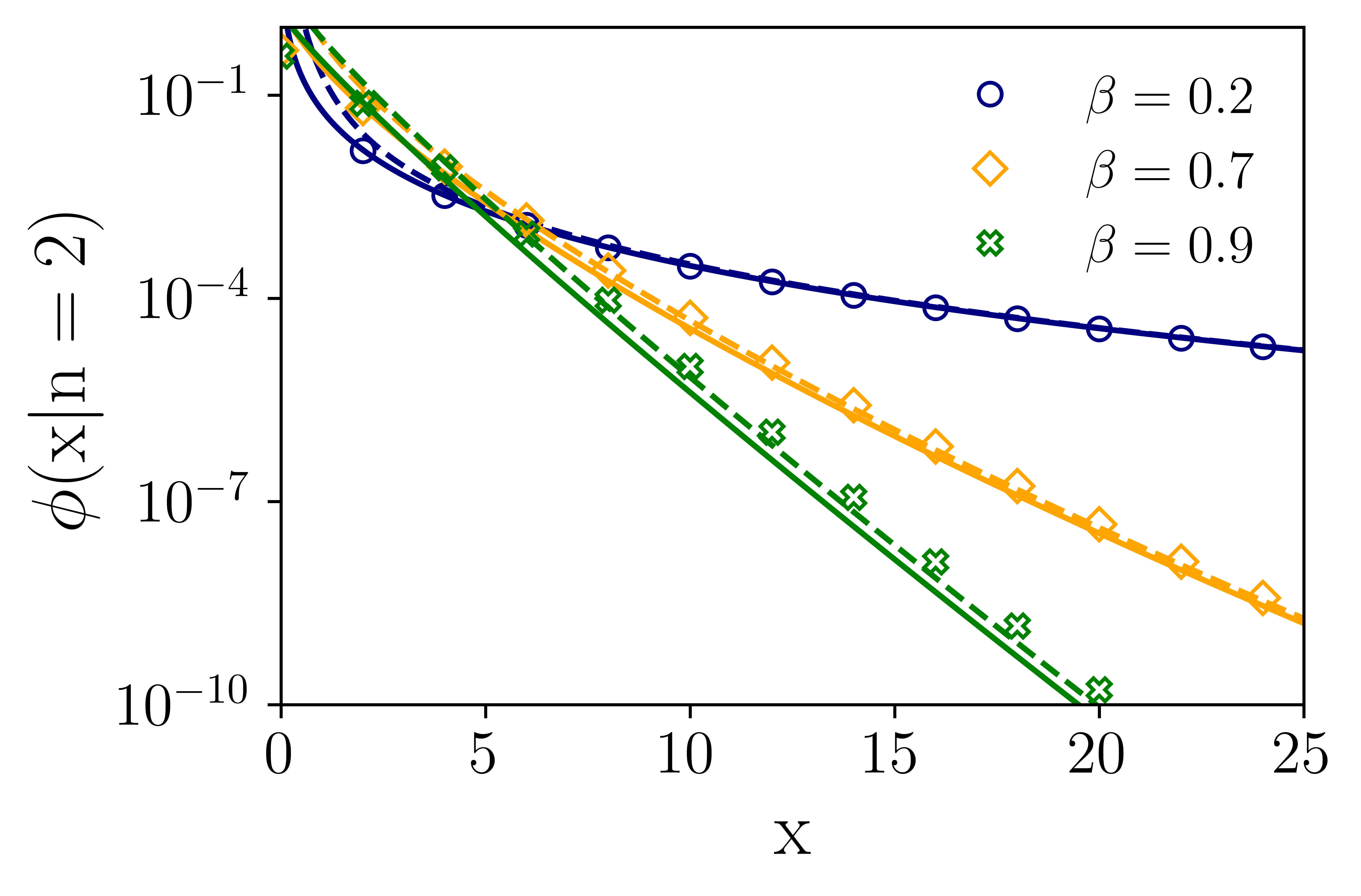}
\caption{The big jump principle for the pedagogical case of two jumps. The figure shows the numerically calculated $\phi(x|2)$ (colored symbols), the leading approximation $(2f(x))$ (colored lines), and the corrected form, derived in Eq. (\ref{eq:BJP n=2 corrected}) (colored dashed lines) for $\beta=0.2,0.7,0.9$, are displayed on a semi-logarithmic scale. While the BJP is effective at the far tails, it becomes apparent that convergence slows as $\beta$ nears the critical value of 1.}\label{Fig:BJP_n=2}
\end{figure}

Now, we calculate corrections to the BJP which are particularly important when $\beta \to 1$. The improved approximation is given by $g(y) \approx -1-|y|^\beta+\beta y +\frac{1}{2}\beta(1-\beta)y^2$, which we can plug into Eq. (\ref{eq:phi(x|2)_exact}) to obtain
\begin{eqnarray}\label{eq:BJP n=2 corrected start}
    &&\phi(x|2) \approx  2\widetilde{N}^2 |x| \nonumber\\
    && \times \int_{-\infty}^{\infty} {\rm e}^{\alpha^\beta|x|^\beta(-1-|y|^\beta)}\big(1+{\rm d}_1(x)y+{\rm d}_2(x)y^2\big)dy,\nonumber\\
\end{eqnarray}
where ${\rm d}_1(x) = \alpha^\beta|x|^\beta\beta$ vanishes due to symmetry. The solution of Eq. (\ref{eq:BJP n=2 corrected start}) yields
\begin{equation}\label{eq:BJP n=2 corrected}
    \phi(x|2) \approx  2\left(1+ \frac{{\rm d}_2(x)}{x^2} \right)f(x),
    \end{equation}
    where
    \begin{equation}
    {\rm d}_2(x) = \frac{1}{2}\alpha^\beta|x|^\beta\beta\left(1-\beta+ \alpha^\beta|x|^\beta \beta\right).
\end{equation}
The first term of Eq. (\ref{eq:BJP n=2 corrected}) is the BJP, as detailed in Eq. (\ref{eq:BJP_leading_term}), while ${\rm d}_2(x)/x^2$, serves as the correction term.
The correction becomes significant as $\beta$ approaches $1$ from below, since the magnitude of ${\rm d}_2(x)/x^2$ is of the same order as the leading term of the BJP.
In Fig. \ref{Fig:BJP_n=2}, we compare the numerically obtained $\phi(x|2)$, the leading term from Eq. (\ref{eq:BJP_leading_term}), and the correction term we have formulated in Eq. (\ref{eq:BJP n=2 corrected}).
This comparison emphasizes the importance of additional correction terms as $\beta \to 1$ and demonstrates their redundancy for $\beta \ll 1$.
\subsection{BJP for CTRW}
As we progress with the big jump principle, integrating Eq. (\ref{eq:BJP_leading_term}) into Eq. (\ref{eq:P(x,t)_summation}) leads us to a straightforward relation \cite{singh2023universal}:
\begin{eqnarray}\label{eq:BJP_P(x,t)_<n>}
    P(x,t) \approx  \langle n_t \rangle f(x),
\end{eqnarray}
where $\langle n_t \rangle$ represents the average number of jumps at time $t$. This relationship holds true for every distribution of waiting times. In this manuscript, our focus is specifically on exponential waiting times, corresponding to Poisson statistics. Namely,  we have $\langle n\rangle =t$, hence $P(x,t)\approx t f(x)$. This signifies a pivotal behavior change: instead of exhibiting exponential tails, the propagator will now display sub-exponential characteristics. Thus, the universality observed in the super-exponential case (for $\beta>1$) disappears for $\beta <1$.
Similarly, if $f(x)$ has a power law tail, Eq. (\ref{eq:BJP_P(x,t)_<n>}) is still valid, and $P(x,t)$ will decay as a power law for large x.
In Fig. \ref{Fig:P(x,t)_BJP_to_LDT}, we showcase this transition, observing how the propagator, $P(x,t)$, shifts from sub-exponential features to exponential, Laplace-like characteristics.
\begin{figure}[ht]
\centering
\includegraphics[width=0.45\textwidth]{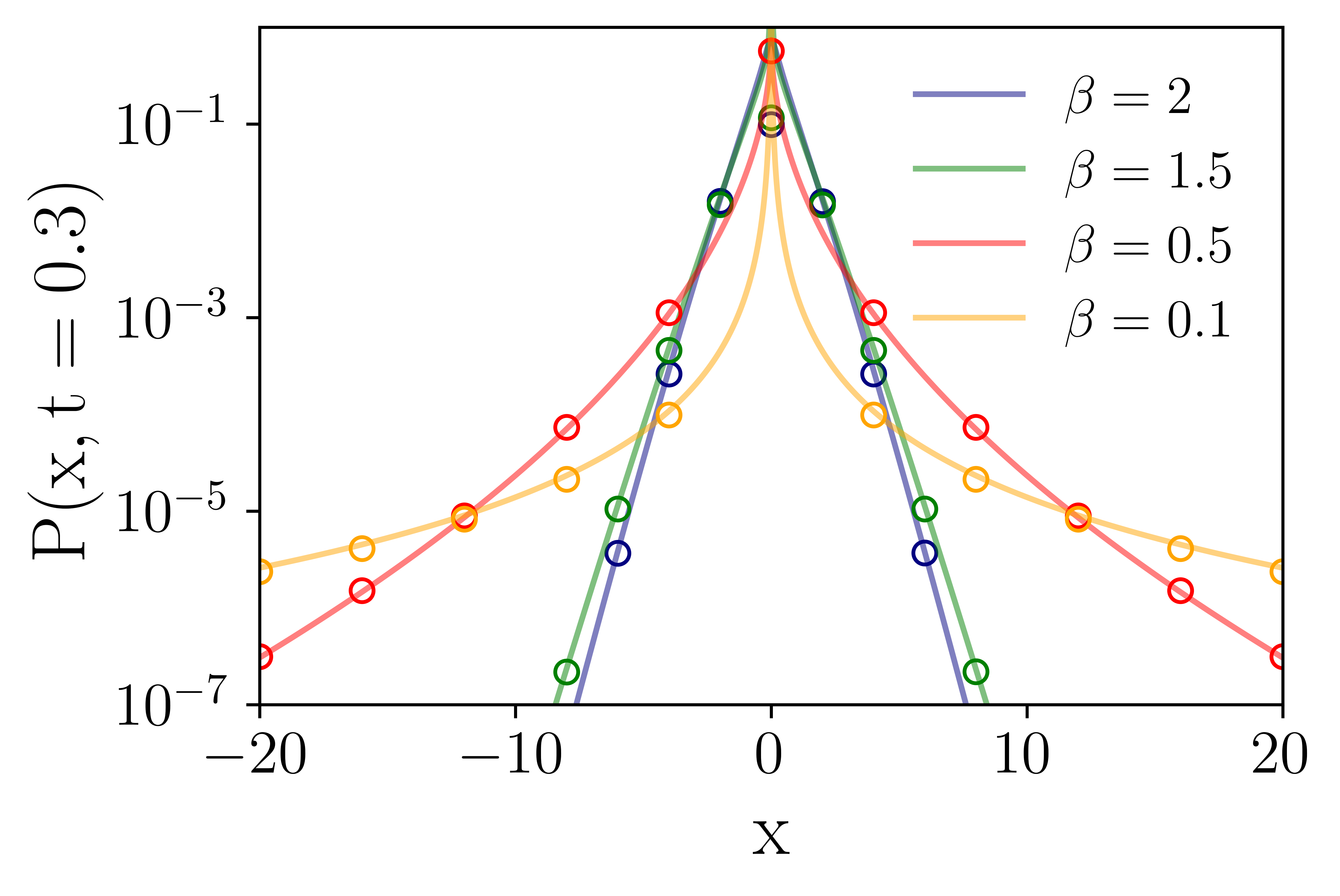}
\caption{The propagator $P(x,t)$ is shown for different values of $\beta$ (shown in legend). The transition of $P(x, t)$ from exponential to Laplace-like characteristics is associated with the large deviations theory. When $ \beta > 1$, we observe a transition to sub-exponential behavior as per the big jump principle when $\beta < 1$. Circles denote the numerically obtained CTRW data, while solid lines represent our approximations. For $ \beta < 1 $, Eq. (\ref{eq:BJP_P(x,t)_<n>}) is employed, and for $ \beta > 1 $, Eq. (\ref{eq:P(x,t)_ARF}) is utilized.
}\label{Fig:P(x,t)_BJP_to_LDT}
\end{figure}

\section{Discussion}
This research has centered on applying large deviations theory (LDT) to the study of diffusive motion in the context of continuous time random walks with exponential waiting times. A notable aspect of LDT is its traditional use in probing rare events, specifically within the large $x$ limit of the probability density function $P(x,t)$. While rate functions, a key component in LDT, are often perceived as challenging to sample in empirical settings \cite{debiossac2023convergence,Thapa_2021}, our study suggests otherwise. By focusing on short time scales and large $x$ values, we demonstrate that the nearly exponential decay of $P(x,t)$ - albeit with logarithmic corrections - is more accessible for sampling than previously assumed.

This accessibility is particularly pertinent in single tracer experiments, where the microscopic time scale, characterized by the mean time between jumps, is not significantly smaller than the observation period $t$. Under these conditions, a universal exponential decay in the particle density is observed. Additionally, our investigation delved into the realm of the big jump principle, particularly its interplay and transition to LDT. The BJP is critical for understanding heavy-tailed processes in which significant fluctuations, or ``big jumps", dominate the characteristics of the random walk. This principle becomes particularly relevant when the jump length distribution is sub-exponential, leading to sub-exponential behavior in the propagator \cite{foss2011introduction,singh2023universal,embrechts2013modelling,kutner2002extreme,de2013asymmetric,majumdar2005nature}. The transition from the BJP regime $(\beta<1)$ to the LDT regime $(\beta>1)$ highlights how different statistical frameworks govern the behavior of diffusive processes under varying conditions. These findings are crucial as they extend the utility of LDT and BJP beyond theoretical confines, making them practical tools for capturing statistical behaviors of particle dispersion that deviate from the normal distribution, inherent in various physical systems.

While the CLT remains a cornerstone in statistical physics, our study illustrates that deviations from CLT predictions, particularly in the tails, are theoretically significant and experimentally observable. Furthermore, the employment of the Edgeworth expansion has provided valuable corrections to the CLT, leading to a more nuanced understanding of the propagator's behavior in different regimes. This approach has bridged the gap between the Gaussian central, and the non-Gaussian tails of the distribution, capturing the essence of diffusive dynamics more comprehensively.

The understanding of Laplace's first law is not limited to the CTRW framework. 
In Refs. \cite{hidalgo2020hitchhiker,baldovin2019polymerization}, a many-body scenario was studied, employing mainly simulations. One example is the Hitchhiker model \cite{hidalgo2020hitchhiker}, a framework based on experimental findings that the size of molecules may fluctuate due to diffusion, aggregation, and fragmentation processes, resulting in exponential-like tails for $P(x,t)$. Similarly, Ref. \cite{baldovin2019polymerization} argues that polymerization processes lead to the demonstrations of Brownian yet non-Gaussian diffusion.

\subsection{Open Questions}
In the CTRW framework, a significant challenge is presented by the intermediate regime, where neither the central limit theorem nor the large deviations theory/big jump principle effectively approximates the results. Rusciano et al.\cite{rusciano2022fickian} introduced a scaling parameter, assuming exponential statistics; $\exp{(-|x|/\lambda(t))}$, where $\lambda(t) \sim t^{1/3}$. An opposing comment was quickly raised by Berthier et al.\cite{berthier2023comment}, yet the phenomenon, particularly the non-trivial exponent $1/3$, was observed in several other experiments \cite{hu2023triggering,aaberg2021glass,rusciano2022fickian,miotto2021length}. More specifically, in Ref. \cite{hu2023triggering}, both the scaling parameter and the Lambert functions, as derived in this manuscript, are demonstrated. Furthermore, in Ref. \cite{wang2009anomalous}, a similar scaling parameter, $\lambda(t) \sim t^{1/2}$, is proposed, adding further intrigue to the subject. It remains unclear whether the experimental and numerical evidence for $\lambda(t) \sim t^b$ represents merely a fitting issue due to insufficient data, or if it describes some deep physics not fully understood by the authors of this manuscript.

Chaudhuri et al. \cite{chaudhuri2007universal} proposed that the waiting time PDF is a combination of two exponentials, with the first jump in the random walk being drawn from a less frequent PDF compared to the subsequent jumps. Under certain conditions, this results in more pronounced exponential tails.
As mentioned in our model, for the sake of mathematical simplicity and broad applicability, we employed a Poisson process, while other works investigated a wider variety of processes \cite{wang2020large}.
However, a practical question that remains largely unanswered is the impact of non-exponential waiting times on the observations made in this study. Similarly, the effect of changing the dimensionality of the problem on our results is yet to be explored.

Another interesting approach to the tails of CTRW was recently pioneered by Sokolov and Pacheco \cite{PhysRevE.103.042116}.
They use the rate function representation of the temporal process (waiting times) and the rate function representation of the spatial process (jumps) to study the properties of the rate function of $P(x,t)$.

Finally, delving beyond sub- and super-exponential jump lengths, we still expect Laplace-like tails, for example, the case of CTRW on a lattice \cite{wang2020large}. One may find similar results to those presented in this manuscript, though the Lambert function used extensively here, is not found to be generic. This implies that the characterization of the rate function for more general jump length distributions might yield further insights. 
\\\\
{\bf Acknowledgements:}
We thank Lucianno Defaveri for insightful discussions. We also acknowledge the support of Israel Science Foundation's grants 1614/21, and 2796/20.
\footnote{%
\begin{tabular}{@{}l@{}} 
* OmerHamdilf2@gmail.com \\
** stasbur@gmail.com \\
*** Eli.Barkai@biu.ac.il
\end{tabular}%
}
\bibliographystyle{apsrev4-1}


%

\end{document}